\documentclass{iau}
\usepackage{graphicx}

\title[MDI of HD~184927] 
{Magnetic Doppler Imaging of He-strong star HD~184927}

\author[Ilya Yakunin, Gregg Wade \& MiMeS Collaboration]   
  {I.~Yakunin$^1$,
  G.~Wade$^2$, D.~Bohlender$^3$, O.~Kochukhov$^4$, V. Tsymbal$^5$
     and MiMeS Collaborators\\}
 \affiliation{  $^1$Special Astrophysical Observatory, Nizhniy Arkhyz, Russia 369167\\
   [\affilskip]
  $^2$Department of Physics, Royal Military College of Canada, P.O. Box 17000, Station Forces, Kingston, Ontario, Canada, K7K 7B4\\
  $^3$National Research Council of Canada, Herzberg Institute of Astrophysics, 5071 West Saanich Road, Victoria, BC, V9E 2E7, Canada\\
  $^4$Department of Physics and Astronomy, Uppsala University, SE-751 20, Uppsala, Sweden\\
  $^5$Tavrian National University, Vernadskiys Avenue 4, Simferopol, Crimea, 95007, Ukraine}

\pubyear{2013}
\volume{302}  
\pagerange{??}
\jname{Magnetic fields throughout stellar evolution}
\editors{A.C. Editor, B.D. Editor \& C.E. Editor, eds.}
\begin{document}

\maketitle

\begin{abstract}
We have employed an extensive new timeseries of Stokes I and V spectra obtained with the ESPaDOnS spectropolarimeter at the 3.6-m Canada-France-Hawaii Telescope to investigate the physical parameters, chemical abundance distributions and magnetic field topology of the slowly-rotating He-strong star HD~184927. We infer a rotation period of $9^d.53071\pm0.00120$ from $H\alpha$, $H\beta$, LSD magnetic measurements and EWs of helium lines. We used an extensive NLTE TLUSTY grid along with the SYNSPEC code to model the observed spectra and find a new value of luminosity. In this poster we present the derived physical parameters of the star and the results of Magnetic Doppler Imaging analysis of the Stokes I and V profiles.  Wide wings of helium lines can be described only under the  assumption of the presence of a large, very helium-rich spot.

\end{abstract}

\firstsection 
\section{Observational data}

Stokes V spectra of HD~184927 were obtained with the ESPaDOnS spectopolarimeter at CFHT between 2008 August 20 and 2012 June 27. The resolution of all the spectra is R=65000. The S/N ratio varies, but is typically is about 520. Spectra were reduced using the Upena pipeline feeding the LIBRE-ESPRIT code. Then we applied the LSD procedure to obtain mean Stokes I and V profiles and improve the S/N ratio of our measurements (see Donati J.-F. \textit{et al.}, 1997, MNRAS, 291, 658 for details). Resulting S/N ratio of LSD profiles varies from 1000 to 1700.

We also used medium resolution data obtained at the  DAO 1.8-meter Plaskett telescope with the dimaPol spectropolarimeter. The S/N in Stokes V varies between 250 and 450 at 4970 \AA. Typically 12-18 10-minute sub-exposures were taken and then combined to produce a single measurement.

To check consistency with previously published data, we included magnetic field measurements from Wade et al. (Wade G. \textit{et al.}, 1997, A\&A, 320, 172) in our dataset.


\section{Period refinement and rotation}

To determine rotational period of HD 184927 we used magnetic field data obtained with SAO Zeeman analyzer and UWO photoelectric polarimeter (see Wade et al. 1997) and added dimaPol measurements of $H\beta$ and He 4922 line. $H\beta$ measurements (UWO + dimaPol) only give possible periods of 9.531072 and 9.522540, with the latter barely within the 1-sigma error bar of the 1997 period estimate. The best ephemeris for this set is $JD=2455706.843 + 9.53071\pm0.00120d$.

In order to estimate projected rotational velocity we used grid of NLTE TLUSTY models with fixed Teff, log g and microturbulence and combination of different [Si/H] and vsini parameters to calculate several synthetic spectra for nine SiII-SiIII lines. The code computed equivalent widths for each case and compared it to the observed equivalent widths. For these calculations we adopted projected rotational velocity $v\sin i = 9\pm2$ km/s.

\section{Magnetic field}
The longitudinal magnetic field and null measurements from the ESPaDOnS spectra were computed from each LSD Stokes V and diagnostic null profile, using an integration range from -200 to 200 km/s. The computed longitudinal field varies from -200 to 950 G. We used the Oblique Rotator Model to model the magnetic curve assuming that the field can be described by a centred dipole. The best-fit model has the following parameters: beta=70, Bd=4500, i=26. To explore the consistency of magnetic field measured from different elements we built LSD masks containing lines of Si, O, N, Fe and He. The longitudinal magnetic field was also measured from the Balmer lines Halpha and Hbeta.

\begin{figure}[h]
\begin{minipage}[h]{0.40\linewidth}
 \center{\includegraphics[width=1\linewidth]{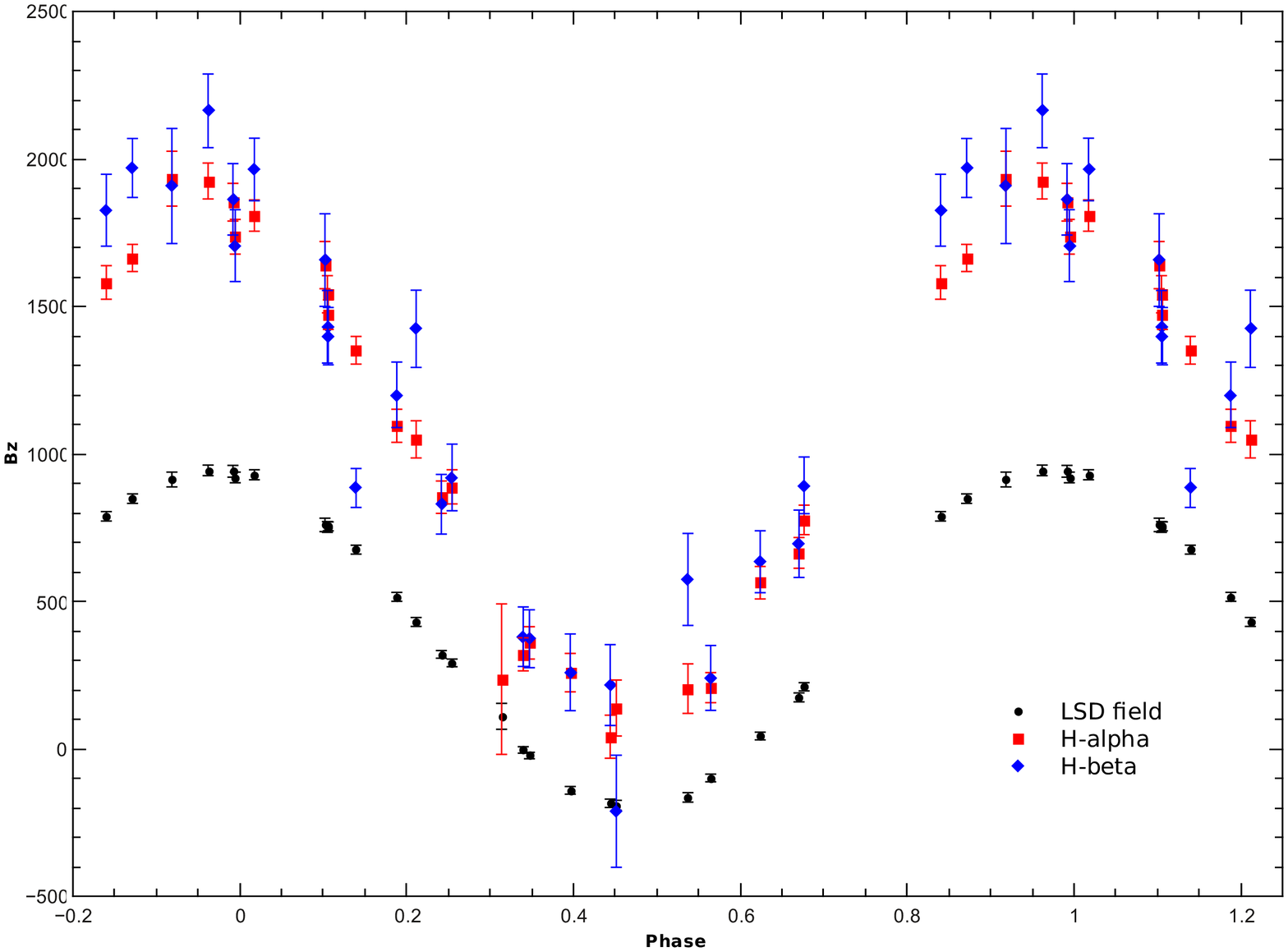}}
 \caption{Longitudinal magnetic field from different spectral lines in comparison with LSD field}
 \label{fig1}
  \end{minipage}
\hfill
\hfill
 \begin{minipage}[h]{0.40\linewidth}
 \center{\includegraphics[width=0.95\linewidth]{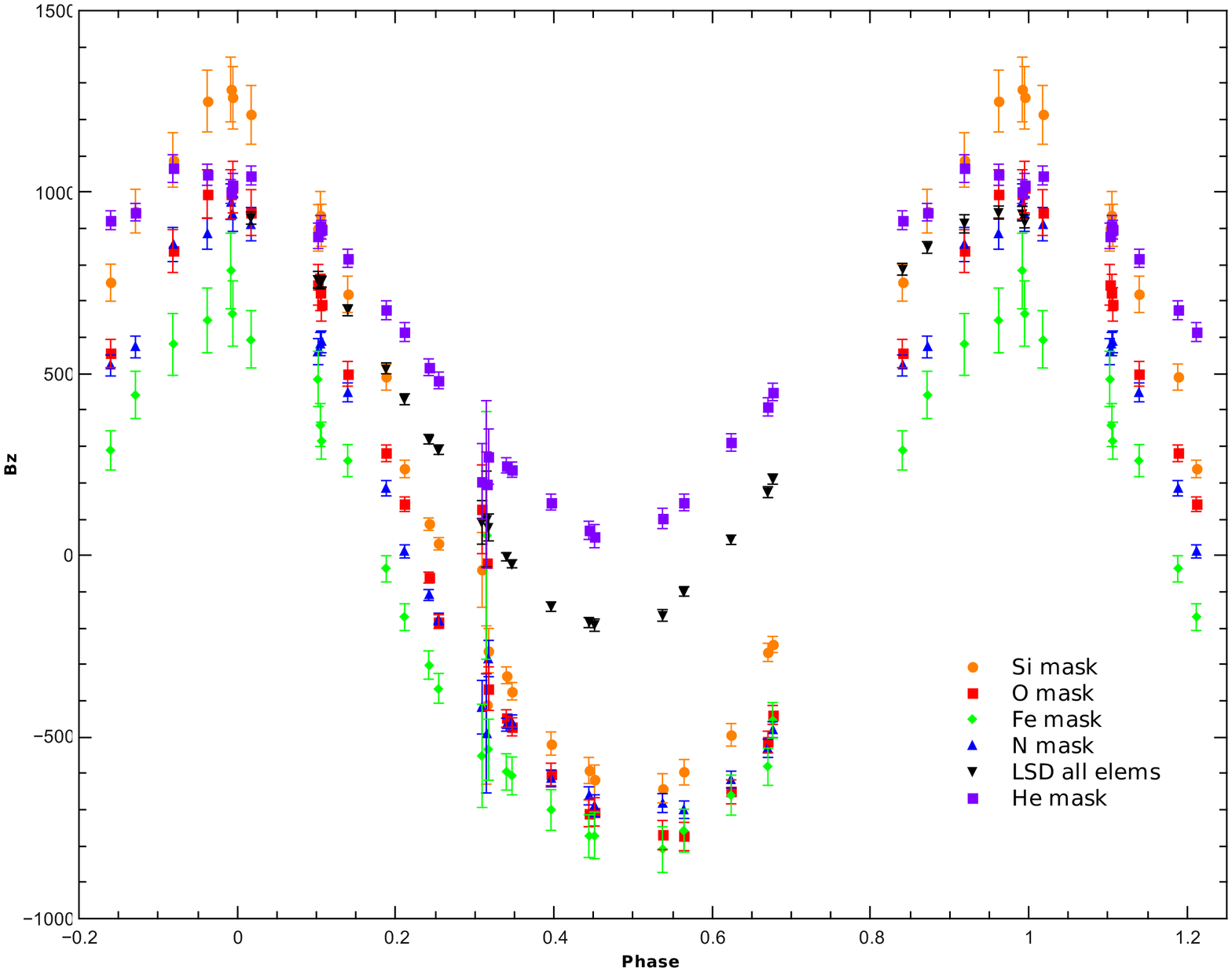}} 
 \caption{Longitudinal magnetic field from different spectral lines in comparison with LSD field}
  \label{fig2}
 \end{minipage}
\end{figure}

As one can clearly see, magnetic field from Si, O, N and Fe masks changes its sign while the field from He is not. At the same time the magnetic field of helium lines strongly correlates with the field measured from hydrogen Halpha and Hbeta lines. We find that the lines can be reproduced by a spot with $60^{\circ}$ radius and helium abundance N(He)/N(H) = 2.

\section{Magnetic Doppler Imaging}
Inversions were carried out with a modified Invers13 code (Kochukhov O. \textit{et al.}, 2013, A\&A, 550, A84). The grid of LLmodels atmospheres (Shulyak D. \textit{et al.}, 2004, A\&A, 428, 993) was computed for Teff=22000, log g=4.0 and a range of He abundance. For this grid we computed NLTE departure coefficients with TLUSTY. These model atmospheres and NLTE departure coefficients were then used in the MDI code to calculate local He I 6678 \AA line profiles. Magnetic field was parameterized with a spherical harmonic expansion, similar to the tau Sco analysis by Donati et al. (Donati J.-F. \textit{et al.}, 2006, MNRAS, 370, 629).

We performed a dipole+quadrupole fitting, allowing full freedom (toroidal field, independent poloidal radial and horizontal fields) of the spherical harmonic expansion. We found that the field is mostly poloidal, with comparable contributions of the dipole and quadrupole terms. Total fraction of poloidal component is 81.5\% and toroidal component is 18.5\%.

\end{document}